\documentclass[prb,preprint,showpacs,preprintnumbers,amsmath,amssymb]{revtex4}

\usepackage{graphicx}
\usepackage{bm}

\newcommand{\be}{\begin{equation}}
\newcommand{\ee}{\end{equation}}
\newcommand{\eq}[1]{Eq.~(\ref{#1})}
\newcommand{\fig}[1]{Fig.~\ref{#1}}
\def\bea{\begin{eqnarray}}
\def\eea{\end{eqnarray}}

\def\bra{\langle}
\def\ket{\rangle}
\def\vq{{\bf q}}

\begin{document}

\title{Charge-Density-Excitation Spectrum in the 
$\boldsymbol{t}$-$\boldsymbol{t'}$-$\boldsymbol{J}$-$\boldsymbol{V}$  Model} 

\author{Andr\'es Greco$^\dag$, Hiroyuki Yamase$^\ddag$, and Mat\'{\i}as Bejas$^\dag$}
\affiliation{
{$^\dag$}Facultad de Ciencias Exactas, Ingenier\'{\i}a y Agrimensura and
Instituto de F\'{\i}sica Rosario (UNR-CONICET),
Av. Pellegrini 250, 2000 Rosario, Argentina\\		
{$^\ddag$}National Institute for Materials Science, Tsukuba 305-0047, Japan
}

\date{August 29, 2016}

\begin{abstract}
We study the density-density correlation function in a large-$N$ scheme of the 
$t$-$t'$-$J$-$V$ model. When the nearest-neighbor Coulomb interaction $V$ is zero, 
our model exhibits phase separation in a wide doping region and we obtain 
large spectral weight near momentum $\vq=(0,0)$ at low energy, which originates from the 
proximity to phase separation. 
These features are much stronger for electron doping than for hole doping.
However, once phase separation is suppressed by including a finite $V$, 
the low-energy spectral weight around $\vq=(0,0)$ is 
substantially suppressed. 
Instead a sharp zero-sound mode is stabilized above the particle-hole continuum. 
We discuss that the presence of a moderate value of $V$, 
which is frequently neglected in the $t$-$J$ model, is important to understand low-energy charge 
excitations especially close to $\vq=(0,0)$ for electron doping. 
This insight should be taken into account in a future study of x-ray scattering measurements. 
\end{abstract}


\maketitle

\section{Introduction}
Charge order and charge excitation spectra attract a renewed interest in cuprate superconductors. 
While the presence of spin-charge stripe order is well known in La-based cuprates \cite{kivelson03}, 
charge order discovered recently is not accompanied by spin order. 
Such charge order is observed not only in hole-doped cuprates  ($h$-cuprates) such as 
Y- \cite{wu11,ghiringhelli12, chang12, achkar12,leboeuf13,blackburn13,blanco-canosa14}, 
Bi- \cite{comin14, da-silva-neto14, hashimoto14}, and Hg-based \cite{tabis14} compounds, 
but also in electron-doped cuprates ($e$-cuprates) \cite{da-silva-neto15}. 

While charge order phenomena are now ubiquitous in cuprates, there is a clear particle-hole 
asymmetry between $h$- and $e$-cuprates. 
Charge order occurs inside the pseudogap phase in $h$-cuprates. In contrast, in $e$-cuprates 
the charge order is observed below $340$K (Ref.~\onlinecite{da-silva-neto15}) and seems 
to occur from the normal metallic phase, since 
the pseudogap effect is absent or very weak in this system.
On the other hand, theoretical insights are controversial with regard to particle-hole 
asymmetry of charge excitations.  
A recent density-matrix renormalization group (DMRG) study 
for the Hubbard model with $U=8t$ 
predicts an enhancement of the low-energy charge excitations in $e$-cuprates, 
whereas a similar enhancement does not occur for $h$-cuprates \cite{tohyama15}. 
In contrast, a leading-order theory of the large-$N$ expansion to the $t$-$J$ model 
shows that a tendency to charge order is stronger in $h$-cuprates 
than $e$-cuprates \cite{bejas14}, implying that low-energy charge excitations tend to 
be enhanced more in $h$-cuprates.  

These different theoretical conclusions could be understood consistently by considering 
the well-known insight that 
$e$-cuprates are expected to be closer to phase separation (PS) than $h$-cuprates 
as shown by various theoretical studies \cite{gooding94,martins01,macridin06,bejas14}. 
However, charge excitations associated with PS are hardly known even theoretically. 
The proximity to PS is expected to have a strong influence on charge excitations especially 
near momentum $\vq=(0,0)$, which will be tested by 
resonant inelastic x-ray scattering (RIXS) measurements in the future. 
Therefore it is important to clarify charge excitations close to PS. 
Moreover, as we will show below, 
PS proves to be a key to resolve the different theoretical insights obtained in 
Refs.~\onlinecite{tohyama15} and \onlinecite{bejas14}. 

In this paper, encouraged by successful explanations of the charge order \cite{da-silva-neto15} and 
the mode near $\vq=(0,0)$ \cite{ishii14,wslee14} in $e$-cuprates in terms of 
a large-$N$ expansion to the $t$-$J$ model \cite{yamase15b,greco16}, 
we compute the density-density correlation function in the same large-$N$ scheme 
as previous ones \cite{yamase15b,greco16}. 
To clarify charge excitations associated with PS, we study the impact of the 
nearest-neighbor Coulomb interaction $V$ and show 
dramatic changes of the low-energy charge excitations around $\vq=(0,0)$ 
especially for electron doping. 
We compare our results with recent theoretical work \cite{tohyama15} and Ref.~\onlinecite{khaliullin96},  
both of which predict the presence of low-energy charge excitations for a small $\vq$.

In Sec.~2 we describe the model and summarize the formalism. Sec.~3 contains the results 
which are discussed in Sec.~4. Conclusions are given in Sec.~5. 

\section{Model and Formalism}
Cuprate superconductors are doped Mott insulators and their minimal model is the so-called 
$t$-$J$ model \cite{anderson87,lee06}. Although the nearest-neighbor Coulomb interaction is 
frequently neglected in a 
study of the $t$-$J$ model, its effect proves to be crucially important to understand charge excitations 
close to $\vq=(0,0)$, as we will show below. Hence we study the following $t$-$t'$-$J$-$V$ model  
\be
H = -\sum_{i, j,\sigma} t_{i j}\tilde{c}^\dag_{i\sigma}\tilde{c}_{j\sigma} + 
J \sum_{\langle i,j \rangle}  \left( \vec{S}_i \cdot \vec{S}_j - \frac{1}{4} n_i n_j \right)
+V \sum_{\bra i, j\ket}  n_i n_j   \,,
\label{tJV}  
\ee
where the sites $i$ and $j$ run over a square lattice. 
The hopping $t_{i j}$ takes a value $t$ $(t')$ between the first (second) nearest-neighbors 
sites.  
$\langle i,j \rangle$ indicates a nearest-neighbor pair, and 
$J$ and $V$ are the spin exchange and Coulomb interaction, respectively. 
$\tilde{c}^\dag_{i\sigma}$ ($\tilde{c}_{i\sigma}$) is 
the creation (annihilation) operator of electrons 
with spin $\sigma$ ($\sigma = \downarrow$,$\uparrow$) 
in the Fock space without double occupancy. 
$n_i=\sum_{\sigma} \tilde{c}^\dag_{i\sigma}\tilde{c}_{i\sigma}$ 
is the electron density operator and $\vec{S}_i$ is the spin operator. 

We analyze the model (\ref{tJV}) in terms of the large-$N$ expansion formulated in Ref.~\onlinecite{foussats04}. 
A full formalism is described in Ref.~\onlinecite{bejas12} where charge instabilities, not charge excitations, 
were studied in the same framework as the present one. Hence leaving the details to Sec.~II~ A in 
Ref.~\onlinecite{bejas12}, we here keep our presentation minimal. 
In the large-$N$ approach, charge excitations with momentum $\vq$ and bosonic 
Matsubara frequency $\omega_{n}$ are described by a $6 \times 6$ bosonic propagator 
\begin{equation}
D^{-1}_{ab}(\vq,\mathrm{i}\omega_n)
= [D^{(0)}_{ab}(\vq,\mathrm{i}\omega_n)]^{-1} - \Pi_{ab}(\vq,\mathrm{i}\omega_n) \,. 
\label{dyson}
\end{equation}
Here $a$ and $b$ run from 1 to 6, 
$D^{(0)}_{ab}(\vq,\mathrm{i}\omega_n)$ is a bare bosonic propagator
\begin{equation} \label{D0inverse}
[D^{(0)}_{ab}(\vq,\mathrm{i}\omega_n)]^{-1} = N 
\left(
\begin{array}{llllll}
F(\vq)
& \frac{\delta}{2} & 0 & 0 & 0 & 0\\
\frac{\delta}{2} & 0 & 0 & 0 & 0 & 0\\
0 & 0 & \frac{4\Delta^2}{J} & 0 & 0 & 0\\
0 & 0 & 0 & \frac{4\Delta^2}{J} & 0 & 0\\
0 & 0 & 0 & 0 & \frac{4\Delta^2}{J} & 0\\
0 & 0 & 0 & 0 & 0 & \frac{4\Delta^2}{J}
\end{array}
\right) ,
\end{equation}
and $\Pi_{ab}(\vq,\mathrm{i}\omega_n)$ are self-energy corrections at leading order; 
$F(\vq)=(\delta^2/2) [ V(\vq)-J(\vq)]$,
$V(\vq)= V  (\cos q_x +  \cos q_y)$ and $J(\vq) = \frac{J}{2} (\cos q_x +  \cos q_y)$; 
for a given doping rate $\delta$, the mean-field value of a bond-field $\Delta$ 
is determined self-consistently.  

We compute the density-density correlation function in the present 
large-$N$ scheme. Summing all contributions up to $O(1/N)$, we obtain 
\begin{eqnarray}
\chi^c({\bf q},{\rm i}\omega_n)=
N {\left( \frac{\delta}{2} \right)}^2 D_{11}({\bf q},{\rm i}\omega_n)\,.
\label{chi}
\end{eqnarray}
Thus, the density-density correlation function is connected with the component $(1,1)$ of the $D_{ab}$.
The factor $N$ in front of Eqs.~(\ref{D0inverse}) and (\ref{chi}) comes from 
the sum over the $N$ fermionic channels after the extension of the spin index 
$\sigma$ from $2$ to $N$. 

Although the physical value is $N=2$, the large-$N$ expansion has several 
advantages over usual perturbation theories. 
First, charge degrees of freedom are generated by carrier doping into a Mott insulator and thus 
the effect of the Coulomb interaction should vanish at half-filling. 
This feature is not reproduced in weak coupling theory\cite{hoang02} but 
in the present leading order theory; see the $(1,1)$ component in \eq{D0inverse}. 
Second, the large-$N$ expansion yields results consistent with those obtained 
by exact diagonalization for charge excitations \cite{merino03} including 
plasmons \cite{prelovsek99,greco16}. 
As we will discuss later, the present formalism also predicts 
charge excitations similar to those obtained in the dynamical DMRG 
method  \cite{tohyama15}. Third, we actually showed that 
the present large-$N$ framework can capture short-range charge order recently observed 
by resonant x-ray scattering \cite{da-silva-neto15, yamase15b} and also a mysterious mode 
around $\vq=(0,0)$ observed by RIXS for $e$-cuprates \cite{ishii14,wslee14,greco16}. 

\section{Results}
As it is well known \cite{gooding94,martins01,macridin06,bejas14}, 
a tendency toward PS is stronger for $e$-cuprates than $h$-cuprates. 
We therefore choose parameters appropriate to $e$-cuprates \cite{yamase15b} such as 
$J/t=0.3$ and $t'/t=0.3$. 
Hereafter all quantities with dimension of energy are measured in units of $t$.  
We compute Im$\chi^{c}(\vq,\omega)$ after analytical continuation in \eq{chi} 
\begin{equation}
{\rm i} \omega_n \rightarrow \omega + {\rm i} \Gamma \,,
\label{gamma}
\end{equation}
where $\Gamma (>0)$ is infinitesimally small and we set $\Gamma=10^{-2}$ for numerical convenience. 
Temperature $T$ is fixed at $T=0$. 
Since PS occurs below $\delta_{c}=0.18$ in the present parameters for $V=0$ [see the inset of 
\fig{weight} (a)], we choose the doping $\delta$ as $\delta=0.20$, which is in the paramagnetic phase 
but close to PS. 
\begin{figure}[t]
\centering
\includegraphics[width=8.5cm]{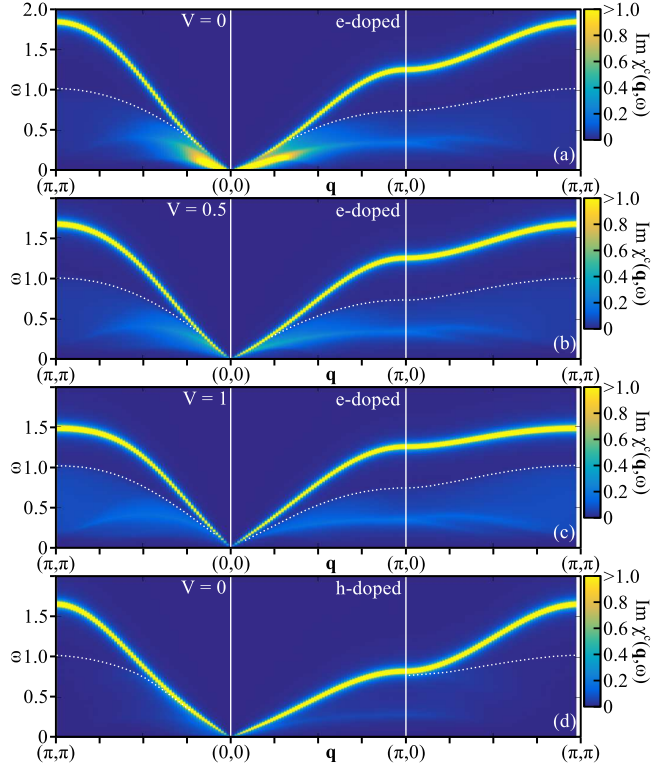}
\caption{(Color online) 
(a)-(c) $\vq$-$\omega$ maps of Im$\chi^{c}(\vq,\omega)$ at $T=0$ and $\delta=0.20$ 
for several choices of Coulomb repulsion $V$. $\vq$ is scanned along the symmetry axes:  
$(\pi,\pi) \rightarrow (0,0) \rightarrow (\pi,0) \rightarrow (\pi,\pi)$.   
(d) $\vq$-$\omega$ map of Im$\chi^{c}(\vq,\omega)$ in the same condition of (a), except 
that a sign of $t'(=-0.3)$ is changed to mimic the hole-doped case.  
}
\label{map}
\end{figure}

Figure~\ref{map} shows intensity maps of Im$\chi^{c}(\vq,\omega)$ for several 
choices of $V$ along the symmetry 
axes: $(\pi,\pi) \rightarrow (0,0) \rightarrow (\pi,0) \rightarrow (\pi,\pi)$. 
The dotted line denotes the upper bound of particle-hole excitations, 
above which there is a sharp dispersive mode. 
This is a particle-hole bound state, called as the zero-sound mode. 
The zero-sound mode was actually obtained in different approximations to the $t$-$J$ model 
\cite{wang91,gehlhoff95,zeyher96a,foussats02,foussats04}. 
Since it is a sound wave, it features a gapless linear dispersion around $\vq=(0,0)$. 
For $V=0$ [\fig{map}(a)], 
there exists large low-energy spectral weight inside the particle-hole continuum slightly below 
the zero-sound mode around $\vq=(0,0)$. 
This low-energy spectral weight diverges upon approaching the PS boundary $\delta_c$, 
leading to the divergence of the compressibility there. Therefore this low-energy spectral weight  
originates from the proximity to PS. 
Because of the mixture of the large low-energy spectral weight, 
the zero-sound mode is overdamped around $\vq=(0,0)$ and becomes less clear. 
By introducing $V$, PS is suppressed as expected [see the inset of \fig{weight} (a)]. 
Concomitantly low-energy spectral 
weight around $\vq=(0,0)$ is also suppressed as shown in Figs.~\ref{map} (b) and (c) for 
$V=0.5$ and $V=1$, respectively. 
We then obtain only the zero-sound mode as dominant charge excitations around $\vq=(0,0)$. 

\begin{figure}[thb]
\centering
\includegraphics[width=7cm]{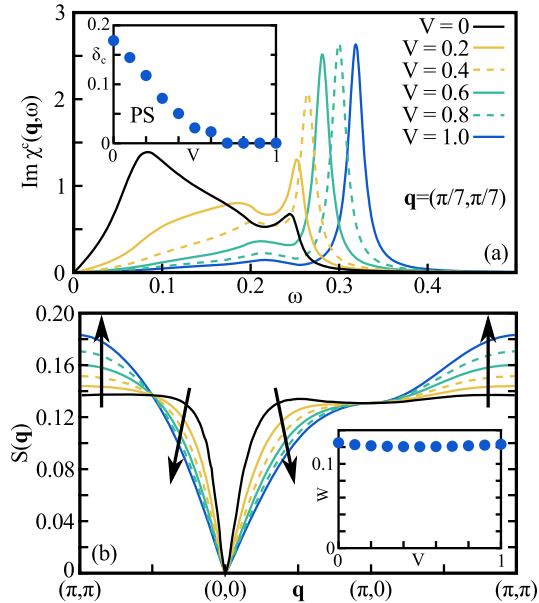}
\caption{(Color online)
(a) $\omega$ dependence of Im$\chi^{c}(\vq,\omega)$ at $\vq=(\pi/7,\pi/7)$ 
and (b) $\vq$ dependence of the integrated spectral weight with respect to $\omega (>0)$ 
for $V=0$, $0.2$, $\cdots$, $1.0$. 
The inset in (a) shows the $V$ dependence of the critical doping, below which PS occurs at $T=0$, 
whereas the inset in (b) the $V$ dependence of the total spectral weight integrated over $\omega (>0)$ 
and $\vq$ along $(\pi,\pi) \rightarrow (0,0) \rightarrow (\pi,0) \rightarrow (\pi,\pi)$. 
}
\label{weight}
\end{figure}

Details of the $V$ dependence of the spectral weight are presented in \fig{weight}(a) 
as a representative, where we present Im$\chi^{c}(\vq,\omega)$ 
at low momentum $\vq=(\pi/7,\pi/7)$ for various choices of $V$.  
With increasing $V$, the spectral weight is transferred to a high energy region 
and the peak associated with the zero-sound mode is enhanced. 
However, when the spectral weight is integrated for each $\vq$ with respect to energy $(>0)$, 
such $\vq$-resolved total spectral weight $S(\vq)$, which corresponds to the equal-time correlation function 
at $T=0$, is suppressed around $\vq=(0,0)$ 
with increasing $V$ as shown in \fig{weight}(b). 
This suggests that the contribution from the proximity to PS is indeed sizable in the 
spectral weight around $\vq=(0,0)$. 
The spectral weight is mainly transferred to a region around $\vq=(\pi,\pi)$. 
This is because a large $V$ favors a checkerboard-type charge-density-wave\cite{foussats04}. 
In fact, such charge order would occur for $V > V_{c} \approx 2.1$, 
but we consider that the region of $V<V_{c}$ is relevant to cuprate superconductors. 
On the other hand, the total weight integrated over $\omega (>0)$ and $\vq$
along the symmetry axes $(\pi,\pi) \rightarrow (0,0) \rightarrow (\pi,0) \rightarrow (\pi,\pi)$ 
does not depend on $V$ [see the inset in \fig{weight}(b)]. 
Therefore the spectral weight 
around $\vq=(0,0)$ is transferred to the region around $\vq=(\pi,\pi)$ with increasing $V$. 
This spectral weight transfer occurs already for V much smaller than $V_c$ and 
we do not observe any charge dynamics, which could be associated with 
other types of charge orders such as stripes from the mechanism of frustrated PS \cite{emery93}. 

For the hole-doped case, we may change the sign of $t'$ \cite{gooding94,tohyama94}. 
In this case, the tendency toward PS becomes much weaker than that for a positive $t'$. 
As seen in \fig{map} (d), 
there is no enhancement of low-energy spectral weight around $\vq=(0,0)$ even for $V=0$. 
In fact,  there occurs no PS for any doping. 
Hence the effect of $V$ becomes much weaker around $\vq=(0,0)$. However, the tendency 
toward charge order at $\vq=(\pi,\pi)$ is common to both sign of $t'$ for a large $V$.

\section{Discussions}
We have found that the nearest-neighbor Coulomb repulsion $V$ plays a crucial role 
to understand the low-energy charge excitations around $\vq=(0,0)$ 
especially in a model calculation for $e$-cuprates, because $e$-cuprates are expected to be close to 
PS \cite{gooding94,martins01,macridin06,bejas14}. 

Recently low-energy charge excitations were reported by a DMRG method 
for a $6 \times 6$ cluster of the Hubbard model with $U=8t$ (Ref.~\onlinecite{tohyama15}). 
In particular, the presence of the low-energy peak was predicted in $e$-cuprates 
at energy lower than the currently available RIXS data \cite{tohyama15}. 

The Hubbard model with a large $U$ should share similar properties to the $t$-$J$ model 
and indeed exhibits phase separation \cite{koch04,macridin06}. 
Hence the predicted low-energy charge excitations likely originate from the proximity to PS and 
would be strongly suppressed once the Coulomb repulsion is included. 
To demonstrate the connection between the present work and Fig.~4 in Ref.~\onlinecite{tohyama15}, 
we took the same momenta, doping, and a large damping $\Gamma=0.1t$ to mimic 
the broadening of Ref.~\onlinecite{tohyama15}; 
$V$ is set to zero. 
We then computed Im$\chi^{c}(\vq,\omega)$ within the present theory for both $t'=0.3t$ (electron-doped case) 
and $t'=-0.3t$ (hole-doped case). Since a large $\Gamma$ was invoked, 
PS does not occur even for the electron-doped case in our framework, but 
charge fluctuations associated with the proximity to PS are expected.

\begin{figure}[bh]
\centering
\includegraphics[width=8.2cm]{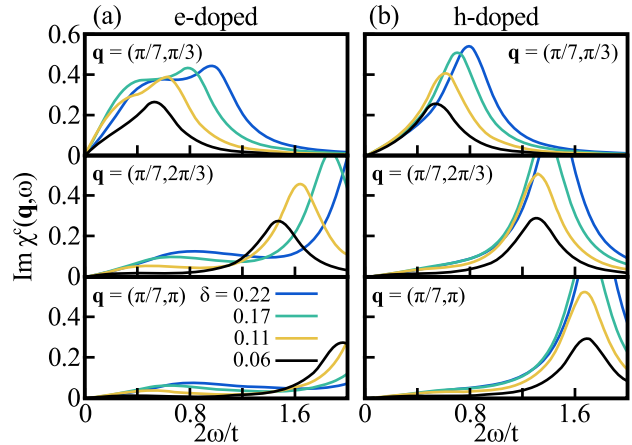}
\caption{(Color online) 
$\omega$ dependence of Im$\chi^{c}(\vq,\omega)$ at $\vq=(\pi/7,\pi/3)$, $(\pi/7,2\pi/3)$, 
and $(\pi/7,\pi)$ for several choices of doping $\delta$ for the electron-doped case (a) 
and the hole-doped case (b). To make a direct comparison with Fig.~4 in Ref.~\onlinecite{tohyama15}, 
$V$ is set to zero, a value of $\Gamma$ is chosen to be $\Gamma=0.1t$, 
and the horizontal axis is taken to be $2 \omega$ 
because $t$ is scaled as $\frac{t}{2}$ in the large-$N$ theory.
}
\label{comparison}
\end{figure}
Figures~\ref{comparison} (a) and (b) capture major features of Figs.~4(c) and (d) in 
Ref.~\onlinecite{tohyama15}, respectively. 
The low-energy spectral weight is strongly suppressed 
at $\vq=(\pi/7,2\pi/3)$ and $(\pi/7,\pi)$ in \fig{comparison}(a),  but 
there is strong enhancement in a low-energy region at $\vq=(\pi/7,\pi/3)$. 
In the hole-doped case [\fig{comparison}(b)] 
the low-energy spectral weight is substantially suppressed 
at $\vq=(\pi/7, 2\pi/3)$ and $(\pi/7,\pi)$, similar to the corresponding results in the electron-doped 
case [\fig{comparison}(a)]. 
At small momentum $\vq=(\pi/7,\pi/3)$, low-energy spectral weight is enhanced more than at 
$\vq=(\pi/7, 2\pi/3)$ and $(\pi/7,\pi)$, 
because of the presence of zero-sound mode at relatively low energy [see \fig{map}(d)]. 
In a lower energy region ($2\omega/t \lesssim 0.4$), we just see the tail of the zero-sound mode and 
do not observe the enhancement of the spectral weight as that seen in 
the electron-doped case [\fig{comparison}(a)].  
This contrast of low-energy charge excitations between 
the electron- and hole-doped case originates from the proximity to PS in the former case. 
In fact, once $V$ is introduced, such low-energy spectral weight is substantially suppressed and both  
electron- and hole-doped cases show similar charge excitations at low energy 
for a small $\vq$, as we have already seen in \fig{map}. 
Low-energy charge excitations at small $\vq$ are also reported in the $t$-$J$ model with $t'=0$ and $V=0$ 
in Ref.~\onlinecite{khaliullin96}. Such excitations may also come from the proximity to PS. 

The zero-sound mode is stabilized in the present model, consistent with the literature 
\cite{wang91,gehlhoff95,zeyher96a,foussats02,foussats04}. We have found that the presence of the zero-sound mode 
is independent of the value of $V$. 
Recalling our parameters $J=0.3$ and $0 \leq V \leq 1.0$, the effective interaction 
$V(\vq)-J(\vq)$ in the $(1,1)$ component of \eq{D0inverse} changes a sign at $V=0.15$. 
Moreover because of the form factor, $\cos q_x + \cos q_y$, the sign of $V(\vq)-J(\vq)$ 
depends on $\vq$. Nevertheless, the zero-sound mode is stabilized for any $V$ 
and $\vq$ in \fig{map}. In fact, as shown in 
Refs.~\onlinecite{gehlhoff95} and \onlinecite{foussats02}, the zero-sound exists even for $J=V=0$. 
This cannot be understood in terms of 
a usual weak coupling analysis such as a random phase approximation (RPA). 

Instead, the robustness of the zero-sound mode originates from strong correlation effects 
contained in the $t$-$J$ model in the sense that the present leading order theory 
does not predict a RPA-like susceptibility, but predicts 
\begin{equation}
\chi^{c}(\vq,\omega) \sim \frac{\Pi_{22}(\vq,\omega)}{[2F(\vq) -\Pi_{11}(\vq,\omega)] 
\Pi_{22} (\vq,\omega) + [\delta -\Pi_{12} (\vq,\omega)]^2} \,.  
\label{chiqw}
\end{equation}
Here we have neglected the components $a,b=3,4,5,6$ in 
Eqs.~(\ref{dyson}) and (\ref{D0inverse}), and their contributions are actually minor.  
We can check that the dispersion of the zero-sound mode is determined by the condition 
\begin{equation}
\left[2F(\vq) - {\rm Re} \Pi_{11}(\vq,\omega) \right] 
{\rm Re} \Pi_{22} (\vq,\omega) + \left[\delta -{\rm Re} \Pi_{12} (\vq,\omega) \right]^2 =0 \,.
\label{mode} 
\end{equation}
The point is that the term, $2F(\vq) - {\rm Re} \Pi_{11}(\vq,\omega)$, 
is always positive in the parameter space we have studied and ${\rm Re} \Pi_{22} (\vq,\omega)$ 
is negative at high energy.  
Equation~(\ref{mode}) is then fulfilled. 
This is the reason why the zero-sound mode is robust in the present model. 
We may interpret high-energy peaks of the density-density 
correlation function obtained by exact diagonalization\cite{tohyama95} 
as the zero-sound mode, as pointed out in Refs.~\onlinecite{gehlhoff95} and \onlinecite{zeyher96a}. 

The robustness of the zero-sound mode, however, should be taken carefully. 
In fact the zero-sound mode changes to a plasmon mode with an excitation gap at $\vq=(0,0)$ 
by including the long-range Coulomb interaction, as recently shown in the present 
large-$N$ scheme for a layered system \cite{greco16}. 

The strong asymmetry of PS between $e$-cuprates ($t'>0$) and $h$-cuprates  ($t'<0$) 
is also understood from \eq{chiqw}. To simplify our analysis, we put $V=J=0$, i.e., $F(\vq)=0$. 
We then obtain at $T=0$ 
\be
\chi^{c}(\vq \rightarrow {\mathbf 0},\omega=0) \sim  \frac{N_{F}}{\delta-4 \mu N_{F}}\,,
\ee
where $N_{F}$ is the density of states at the Fermi energy 
and $\mu$ is the chemical potential \cite{bejas12}. 
Hence PS occurs when the following condition is fulfilled:  
\be
\delta < 4 \mu N_{F} \,.
\label{PScondition}
\ee
We can check that $\mu$ becomes positive close to half-filling only for $t'>0$ 
and \eq{PScondition} is fulfilled, 
showing that PS can occur only for the electron-doped case. 
We can also check that \eq{PScondition} indeed explains our previous numerical results, 
Fig.~3(a) for $V=J=0$ in Ref.~\onlinecite{bejas14}. 
Therefore the asymmetry of PS between $e$- and $h$-cuprates is controlled 
by \eq{PScondition} and originates from 
strong correlation effects in the sense that \eq{PScondition} is obtained for $V=J=0$. 
For a finite $J$ and $V$ we can perform a similar analysis and obtain the same conclusion.

\section{Conclusions} 
We have studied charge excitations in the $t$-$t'$-$J$ model by including 
the nearest-neighbor Coulomb repulsion $V$. 
While the effect of $V$ is frequently neglected in research of charge excitations in cuprates, 
we have found that the $V$ term is crucially important to understand low-energy charge excitations around $\vq=(0,0)$ 
[Figs.~\ref{map}(a)-(c) and \ref{weight}(a)] especially in a model calculation for $e$-cuprates. 
In line with the prediction in Ref.~\onlinecite{tohyama15}, we have also obtained 
low-energy spectral weight around $\vq=(0,0)$, 
but this comes from the proximity to PS and holds only if the effect of $V$ is sufficiently small. 
Given the increasing interest in the study of low-energy charge 
excitations around $\vq=(0,0)$ in RIXS, it is crucial to consider the presence of 
the Coulomb repulsion, which is expected to be finite in real systems. 
We have also found that the zero-sound mode is stabilized above 
the particle-hole continuum, which is independent of a value of $V$. 

\acknowledgments
The authors thank G. Khaliullin and T. Tohyama for very fruitful discussions. 
H.Y. acknowledges support by JSPS KAKENHI Grant Number 15K05189.

\bibliography{main} 

\begin{thebibliography}{38}
\expandafter\ifx\csname natexlab\endcsname\relax\def\natexlab#1{#1}\fi
\expandafter\ifx\csname bibnamefont\endcsname\relax
  \def\bibnamefont#1{#1}\fi
\expandafter\ifx\csname bibfnamefont\endcsname\relax
  \def\bibfnamefont#1{#1}\fi
\expandafter\ifx\csname citenamefont\endcsname\relax
  \def\citenamefont#1{#1}\fi
\expandafter\ifx\csname url\endcsname\relax
  \def\url#1{\texttt{#1}}\fi
\expandafter\ifx\csname urlprefix\endcsname\relax\def\urlprefix{URL }\fi
\providecommand{\bibinfo}[2]{#2}
\providecommand{\eprint}[2][]{\url{#2}}

\bibitem[{\citenamefont{Kivelson et~al.}(2003)\citenamefont{Kivelson, Bindloss,
  Fradkin, Oganesyan, Tranquada, Kapitulnik, and Howald}}]{kivelson03}
\bibinfo{author}{\bibfnamefont{S.~A.} \bibnamefont{Kivelson}},
  \bibinfo{author}{\bibfnamefont{I.~P.} \bibnamefont{Bindloss}},
  \bibinfo{author}{\bibfnamefont{E.}~\bibnamefont{Fradkin}},
  \bibinfo{author}{\bibfnamefont{V.}~\bibnamefont{Oganesyan}},
  \bibinfo{author}{\bibfnamefont{J.~M.} \bibnamefont{Tranquada}},
  \bibinfo{author}{\bibfnamefont{A.}~\bibnamefont{Kapitulnik}},
  \bibnamefont{and} \bibinfo{author}{\bibfnamefont{C.}~\bibnamefont{Howald}},
  \bibinfo{journal}{Rev. Mod. Phys.} \textbf{\bibinfo{volume}{75}},
  \bibinfo{pages}{1201} (\bibinfo{year}{2003}).

\bibitem[{\citenamefont{Wu et~al.}(2011)\citenamefont{Wu, Mayaffre, Kr{\"a}mer,
  Horvati{\'c}, Berthier, Hardy, Liang, Bon, and Julien}}]{wu11}
\bibinfo{author}{\bibfnamefont{T.}~\bibnamefont{Wu}},
  \bibinfo{author}{\bibfnamefont{H.}~\bibnamefont{Mayaffre}},
  \bibinfo{author}{\bibfnamefont{S.}~\bibnamefont{Kr{\"a}mer}},
  \bibinfo{author}{\bibfnamefont{M.}~\bibnamefont{Horvati{\'c}}},
  \bibinfo{author}{\bibfnamefont{C.}~\bibnamefont{Berthier}},
  \bibinfo{author}{\bibfnamefont{W.~N.} \bibnamefont{Hardy}},
  \bibinfo{author}{\bibfnamefont{R.}~\bibnamefont{Liang}},
  \bibinfo{author}{\bibfnamefont{D.~A.} \bibnamefont{Bon}}, \bibnamefont{and}
  \bibinfo{author}{\bibfnamefont{M.-H.} \bibnamefont{Julien}},
  \bibinfo{journal}{Nature} \textbf{\bibinfo{volume}{477}},
  \bibinfo{pages}{191} (\bibinfo{year}{2011}).

\bibitem[{\citenamefont{Ghiringhelli et~al.}(2012)\citenamefont{Ghiringhelli,
  Tacon, Minola, Blanco-Canosa, Mazzoli, Brookes, Luca, Frano, Hawthorn, He
  et~al.}}]{ghiringhelli12}
\bibinfo{author}{\bibfnamefont{G.}~\bibnamefont{Ghiringhelli}},
  \bibinfo{author}{\bibfnamefont{M.~L.} \bibnamefont{Tacon}},
  \bibinfo{author}{\bibfnamefont{M.}~\bibnamefont{Minola}},
  \bibinfo{author}{\bibfnamefont{S.}~\bibnamefont{Blanco-Canosa}},
  \bibinfo{author}{\bibfnamefont{C.}~\bibnamefont{Mazzoli}},
  \bibinfo{author}{\bibfnamefont{N.~B.} \bibnamefont{Brookes}},
  \bibinfo{author}{\bibfnamefont{G.~M.~D.} \bibnamefont{Luca}},
  \bibinfo{author}{\bibfnamefont{A.}~\bibnamefont{Frano}},
  \bibinfo{author}{\bibfnamefont{D.~G.} \bibnamefont{Hawthorn}},
  \bibinfo{author}{\bibfnamefont{F.}~\bibnamefont{He}}, \bibnamefont{et~al.},
  \bibinfo{journal}{Science} \textbf{\bibinfo{volume}{337}},
  \bibinfo{pages}{821} (\bibinfo{year}{2012}).

\bibitem[{\citenamefont{Chang et~al.}(2012)\citenamefont{Chang, Blackburn,
  Holmes, Christensen, Larsen, Mesot, Liang, Bonn, Hardy, Watenphul
  et~al.}}]{chang12}
\bibinfo{author}{\bibfnamefont{J.}~\bibnamefont{Chang}},
  \bibinfo{author}{\bibfnamefont{E.}~\bibnamefont{Blackburn}},
  \bibinfo{author}{\bibfnamefont{A.~T.} \bibnamefont{Holmes}},
  \bibinfo{author}{\bibfnamefont{N.~B.} \bibnamefont{Christensen}},
  \bibinfo{author}{\bibfnamefont{J.}~\bibnamefont{Larsen}},
  \bibinfo{author}{\bibfnamefont{J.}~\bibnamefont{Mesot}},
  \bibinfo{author}{\bibfnamefont{R.}~\bibnamefont{Liang}},
  \bibinfo{author}{\bibfnamefont{D.~A.} \bibnamefont{Bonn}},
  \bibinfo{author}{\bibfnamefont{W.~N.} \bibnamefont{Hardy}},
  \bibinfo{author}{\bibfnamefont{A.}~\bibnamefont{Watenphul}},
  \bibnamefont{et~al.}, \bibinfo{journal}{Nat. Phys.}
  \textbf{\bibinfo{volume}{8}}, \bibinfo{pages}{871} (\bibinfo{year}{2012}).

\bibitem[{\citenamefont{Achkar et~al.}(2012)\citenamefont{Achkar, Sutarto, Mao,
  He, Frano, Blanco-Canosa, Le~Tacon, Ghiringhelli, Braicovich, Minola
  et~al.}}]{achkar12}
\bibinfo{author}{\bibfnamefont{A.~J.} \bibnamefont{Achkar}},
  \bibinfo{author}{\bibfnamefont{R.}~\bibnamefont{Sutarto}},
  \bibinfo{author}{\bibfnamefont{X.}~\bibnamefont{Mao}},
  \bibinfo{author}{\bibfnamefont{F.}~\bibnamefont{He}},
  \bibinfo{author}{\bibfnamefont{A.}~\bibnamefont{Frano}},
  \bibinfo{author}{\bibfnamefont{S.}~\bibnamefont{Blanco-Canosa}},
  \bibinfo{author}{\bibfnamefont{M.}~\bibnamefont{Le~Tacon}},
  \bibinfo{author}{\bibfnamefont{G.}~\bibnamefont{Ghiringhelli}},
  \bibinfo{author}{\bibfnamefont{L.}~\bibnamefont{Braicovich}},
  \bibinfo{author}{\bibfnamefont{M.}~\bibnamefont{Minola}},
  \bibnamefont{et~al.}, \bibinfo{journal}{Phys. Rev. Lett.}
  \textbf{\bibinfo{volume}{109}}, \bibinfo{pages}{167001}
  (\bibinfo{year}{2012}).

\bibitem[{\citenamefont{LeBoeuf et~al.}(2013)\citenamefont{LeBoeuf, Kr\"amer,
  Hardy, Liang, Bonn, and Proust}}]{leboeuf13}
\bibinfo{author}{\bibfnamefont{D.}~\bibnamefont{LeBoeuf}},
  \bibinfo{author}{\bibfnamefont{S.}~\bibnamefont{Kr\"amer}},
  \bibinfo{author}{\bibfnamefont{W.~N.} \bibnamefont{Hardy}},
  \bibinfo{author}{\bibfnamefont{R.}~\bibnamefont{Liang}},
  \bibinfo{author}{\bibfnamefont{D.~A.} \bibnamefont{Bonn}}, \bibnamefont{and}
  \bibinfo{author}{\bibfnamefont{C.}~\bibnamefont{Proust}},
  \bibinfo{journal}{Nat. Phys.} \textbf{\bibinfo{volume}{9}},
  \bibinfo{pages}{79} (\bibinfo{year}{2013}).

\bibitem[{\citenamefont{Blackburn et~al.}(2013)\citenamefont{Blackburn, Chang,
  H{\"u}cker, Holmes, Christensen, Liang, Bonn, Hardy, R{\"u}tt, Gutowski
  et~al.}}]{blackburn13}
\bibinfo{author}{\bibfnamefont{E.}~\bibnamefont{Blackburn}},
  \bibinfo{author}{\bibfnamefont{J.}~\bibnamefont{Chang}},
  \bibinfo{author}{\bibfnamefont{M.}~\bibnamefont{H{\"u}cker}},
  \bibinfo{author}{\bibfnamefont{A.~T.} \bibnamefont{Holmes}},
  \bibinfo{author}{\bibfnamefont{N.~B.} \bibnamefont{Christensen}},
  \bibinfo{author}{\bibfnamefont{R.}~\bibnamefont{Liang}},
  \bibinfo{author}{\bibfnamefont{D.~A.} \bibnamefont{Bonn}},
  \bibinfo{author}{\bibfnamefont{W.~N.} \bibnamefont{Hardy}},
  \bibinfo{author}{\bibfnamefont{U.}~\bibnamefont{R{\"u}tt}},
  \bibinfo{author}{\bibfnamefont{O.}~\bibnamefont{Gutowski}},
  \bibnamefont{et~al.}, \bibinfo{journal}{Phys. Rev. Lett.}
  \textbf{\bibinfo{volume}{110}}, \bibinfo{pages}{137004}
  (\bibinfo{year}{2013}).

\bibitem[{\citenamefont{Blanco-Canosa et~al.}(2014)\citenamefont{Blanco-Canosa,
  Frano, Schierle, Porras, Loew, Minola, Bluschke, Weschke, Keimer, and
  Tacon}}]{blanco-canosa14}
\bibinfo{author}{\bibfnamefont{S.}~\bibnamefont{Blanco-Canosa}},
  \bibinfo{author}{\bibfnamefont{A.}~\bibnamefont{Frano}},
  \bibinfo{author}{\bibfnamefont{E.}~\bibnamefont{Schierle}},
  \bibinfo{author}{\bibfnamefont{J.}~\bibnamefont{Porras}},
  \bibinfo{author}{\bibfnamefont{T.}~\bibnamefont{Loew}},
  \bibinfo{author}{\bibfnamefont{M.}~\bibnamefont{Minola}},
  \bibinfo{author}{\bibfnamefont{M.}~\bibnamefont{Bluschke}},
  \bibinfo{author}{\bibfnamefont{E.}~\bibnamefont{Weschke}},
  \bibinfo{author}{\bibfnamefont{B.}~\bibnamefont{Keimer}}, \bibnamefont{and}
  \bibinfo{author}{\bibfnamefont{M.~L.} \bibnamefont{Tacon}},
  \bibinfo{journal}{Phys. Rev. B} \textbf{\bibinfo{volume}{90}},
  \bibinfo{pages}{054513} (\bibinfo{year}{2014}).

\bibitem[{\citenamefont{Comin et~al.}(2014)\citenamefont{Comin, Frano, Yee,
  Yoshida, Eisaki, Schierle, Weschke, Sutarto, He, Soumyanarayanan
  et~al.}}]{comin14}
\bibinfo{author}{\bibfnamefont{R.}~\bibnamefont{Comin}},
  \bibinfo{author}{\bibfnamefont{A.}~\bibnamefont{Frano}},
  \bibinfo{author}{\bibfnamefont{M.~M.} \bibnamefont{Yee}},
  \bibinfo{author}{\bibfnamefont{Y.}~\bibnamefont{Yoshida}},
  \bibinfo{author}{\bibfnamefont{H.}~\bibnamefont{Eisaki}},
  \bibinfo{author}{\bibfnamefont{E.}~\bibnamefont{Schierle}},
  \bibinfo{author}{\bibfnamefont{E.}~\bibnamefont{Weschke}},
  \bibinfo{author}{\bibfnamefont{R.}~\bibnamefont{Sutarto}},
  \bibinfo{author}{\bibfnamefont{F.}~\bibnamefont{He}},
  \bibinfo{author}{\bibfnamefont{A.}~\bibnamefont{Soumyanarayanan}},
  \bibnamefont{et~al.}, \bibinfo{journal}{Science}
  \textbf{\bibinfo{volume}{343}}, \bibinfo{pages}{390} (\bibinfo{year}{2014}).

\bibitem[{\citenamefont{da~Silva~Neto et~al.}(2014)\citenamefont{da~Silva~Neto,
  Aynajian, Frano, Comin, Schierle, Weschke, Gyenis, Wen, Schneeloch, Xu
  et~al.}}]{da-silva-neto14}
\bibinfo{author}{\bibfnamefont{E.~H.} \bibnamefont{da~Silva~Neto}},
  \bibinfo{author}{\bibfnamefont{P.}~\bibnamefont{Aynajian}},
  \bibinfo{author}{\bibfnamefont{A.}~\bibnamefont{Frano}},
  \bibinfo{author}{\bibfnamefont{R.}~\bibnamefont{Comin}},
  \bibinfo{author}{\bibfnamefont{E.}~\bibnamefont{Schierle}},
  \bibinfo{author}{\bibfnamefont{E.}~\bibnamefont{Weschke}},
  \bibinfo{author}{\bibfnamefont{A.}~\bibnamefont{Gyenis}},
  \bibinfo{author}{\bibfnamefont{J.}~\bibnamefont{Wen}},
  \bibinfo{author}{\bibfnamefont{J.}~\bibnamefont{Schneeloch}},
  \bibinfo{author}{\bibfnamefont{Z.}~\bibnamefont{Xu}}, \bibnamefont{et~al.},
  \bibinfo{journal}{Science} \textbf{\bibinfo{volume}{343}},
  \bibinfo{pages}{393} (\bibinfo{year}{2014}).

\bibitem[{\citenamefont{Hashimoto et~al.}(2014)\citenamefont{Hashimoto,
  Ghiringhelli, Lee, Dellea, Amorese, Mazzoli, Kummer, Brookes, Moritz, Yoshida
  et~al.}}]{hashimoto14}
\bibinfo{author}{\bibfnamefont{M.}~\bibnamefont{Hashimoto}},
  \bibinfo{author}{\bibfnamefont{G.}~\bibnamefont{Ghiringhelli}},
  \bibinfo{author}{\bibfnamefont{W.-S.} \bibnamefont{Lee}},
  \bibinfo{author}{\bibfnamefont{G.}~\bibnamefont{Dellea}},
  \bibinfo{author}{\bibfnamefont{A.}~\bibnamefont{Amorese}},
  \bibinfo{author}{\bibfnamefont{C.}~\bibnamefont{Mazzoli}},
  \bibinfo{author}{\bibfnamefont{K.}~\bibnamefont{Kummer}},
  \bibinfo{author}{\bibfnamefont{N.~B.} \bibnamefont{Brookes}},
  \bibinfo{author}{\bibfnamefont{B.}~\bibnamefont{Moritz}},
  \bibinfo{author}{\bibfnamefont{Y.}~\bibnamefont{Yoshida}},
  \bibnamefont{et~al.}, \bibinfo{journal}{Phys. Rev. B}
  \textbf{\bibinfo{volume}{89}}, \bibinfo{pages}{220511}
  (\bibinfo{year}{2014}).

\bibitem[{\citenamefont{Tabis et~al.}(2014)\citenamefont{Tabis, Li, Tacon,
  Braicovich, Kreyssig, Minola, Dellea, Weschke, Veit, Ramazanoglu
  et~al.}}]{tabis14}
\bibinfo{author}{\bibfnamefont{W.}~\bibnamefont{Tabis}},
  \bibinfo{author}{\bibfnamefont{Y.}~\bibnamefont{Li}},
  \bibinfo{author}{\bibfnamefont{M.~L.} \bibnamefont{Tacon}},
  \bibinfo{author}{\bibfnamefont{L.}~\bibnamefont{Braicovich}},
  \bibinfo{author}{\bibfnamefont{A.}~\bibnamefont{Kreyssig}},
  \bibinfo{author}{\bibfnamefont{M.}~\bibnamefont{Minola}},
  \bibinfo{author}{\bibfnamefont{G.}~\bibnamefont{Dellea}},
  \bibinfo{author}{\bibfnamefont{E.}~\bibnamefont{Weschke}},
  \bibinfo{author}{\bibfnamefont{M.~J.} \bibnamefont{Veit}},
  \bibinfo{author}{\bibfnamefont{M.}~\bibnamefont{Ramazanoglu}},
  \bibnamefont{et~al.}, \bibinfo{journal}{Nat. Commun.}
  \textbf{\bibinfo{volume}{5}}, \bibinfo{pages}{5875} (\bibinfo{year}{2014}).

\bibitem[{\citenamefont{da~Silva~Neto et~al.}(2015)\citenamefont{da~Silva~Neto,
  Comin, He, Sutarto, Jiang, Greene, Sawatzky, and
  Damascelli}}]{da-silva-neto15}
\bibinfo{author}{\bibfnamefont{E.~H.} \bibnamefont{da~Silva~Neto}},
  \bibinfo{author}{\bibfnamefont{R.}~\bibnamefont{Comin}},
  \bibinfo{author}{\bibfnamefont{F.}~\bibnamefont{He}},
  \bibinfo{author}{\bibfnamefont{R.}~\bibnamefont{Sutarto}},
  \bibinfo{author}{\bibfnamefont{Y.}~\bibnamefont{Jiang}},
  \bibinfo{author}{\bibfnamefont{R.~L.} \bibnamefont{Greene}},
  \bibinfo{author}{\bibfnamefont{G.~A.} \bibnamefont{Sawatzky}},
  \bibnamefont{and}
  \bibinfo{author}{\bibfnamefont{A.}~\bibnamefont{Damascelli}},
  \bibinfo{journal}{Science} \textbf{\bibinfo{volume}{347}},
  \bibinfo{pages}{282} (\bibinfo{year}{2015}).

\bibitem[{\citenamefont{Tohyama et~al.}(2015)\citenamefont{Tohyama, Tsutsui,
  Mori, Sota, and Yunoki}}]{tohyama15}
\bibinfo{author}{\bibfnamefont{T.}~\bibnamefont{Tohyama}},
  \bibinfo{author}{\bibfnamefont{K.}~\bibnamefont{Tsutsui}},
  \bibinfo{author}{\bibfnamefont{M.}~\bibnamefont{Mori}},
  \bibinfo{author}{\bibfnamefont{S.}~\bibnamefont{Sota}}, \bibnamefont{and}
  \bibinfo{author}{\bibfnamefont{S.}~\bibnamefont{Yunoki}},
  \bibinfo{journal}{Phys. \ Rev. \ B} \textbf{\bibinfo{volume}{92}},
  \bibinfo{pages}{014515} (\bibinfo{year}{2015}).

\bibitem[{\citenamefont{Bejas et~al.}(2014)\citenamefont{Bejas, Greco, and
  Yamase}}]{bejas14}
\bibinfo{author}{\bibfnamefont{M.}~\bibnamefont{Bejas}},
  \bibinfo{author}{\bibfnamefont{A.}~\bibnamefont{Greco}}, \bibnamefont{and}
  \bibinfo{author}{\bibfnamefont{H.}~\bibnamefont{Yamase}},
  \bibinfo{journal}{New J. Phys.} \textbf{\bibinfo{volume}{16}},
  \bibinfo{pages}{123002} (\bibinfo{year}{2014}).

\bibitem[{\citenamefont{Gooding et~al.}(1994)\citenamefont{Gooding, Vos, and
  Leung}}]{gooding94}
\bibinfo{author}{\bibfnamefont{R.~J.} \bibnamefont{Gooding}},
  \bibinfo{author}{\bibfnamefont{K.~J.~E.} \bibnamefont{Vos}},
  \bibnamefont{and} \bibinfo{author}{\bibfnamefont{P.~W.} \bibnamefont{Leung}},
  \bibinfo{journal}{Phys.\ Rev.\ B} \textbf{\bibinfo{volume}{50}},
  \bibinfo{pages}{12 866} (\bibinfo{year}{1994}).

\bibitem[{\citenamefont{Martins et~al.}(2001)\citenamefont{Martins, Xavier,
  Arrachea, and Dagotto}}]{martins01}
\bibinfo{author}{\bibfnamefont{G.~B.} \bibnamefont{Martins}},
  \bibinfo{author}{\bibfnamefont{J.~C.} \bibnamefont{Xavier}},
  \bibinfo{author}{\bibfnamefont{L.}~\bibnamefont{Arrachea}}, \bibnamefont{and}
  \bibinfo{author}{\bibfnamefont{E.}~\bibnamefont{Dagotto}},
  \bibinfo{journal}{Phys.\ Rev.\ B} \textbf{\bibinfo{volume}{64}},
  \bibinfo{pages}{R1805} (\bibinfo{year}{2001}).

\bibitem[{\citenamefont{Macridin et~al.}(2006)\citenamefont{Macridin, Jarrell,
  and Maier}}]{macridin06}
\bibinfo{author}{\bibfnamefont{A.}~\bibnamefont{Macridin}},
  \bibinfo{author}{\bibfnamefont{M.}~\bibnamefont{Jarrell}}, \bibnamefont{and}
  \bibinfo{author}{\bibfnamefont{T.}~\bibnamefont{Maier}},
  \bibinfo{journal}{Phys.\ Rev.\ B} \textbf{\bibinfo{volume}{74}},
  \bibinfo{pages}{085104} (\bibinfo{year}{2006}).

\bibitem[{\citenamefont{Ishii et~al.}(2014)\citenamefont{Ishii, Fujita, Sasaki,
  Minola, Dellea, Mazzoli, Kummer, Ghiringhelli, Braicovich, Tohyama
  et~al.}}]{ishii14}
\bibinfo{author}{\bibfnamefont{K.}~\bibnamefont{Ishii}},
  \bibinfo{author}{\bibfnamefont{M.}~\bibnamefont{Fujita}},
  \bibinfo{author}{\bibfnamefont{T.}~\bibnamefont{Sasaki}},
  \bibinfo{author}{\bibfnamefont{M.}~\bibnamefont{Minola}},
  \bibinfo{author}{\bibfnamefont{G.}~\bibnamefont{Dellea}},
  \bibinfo{author}{\bibfnamefont{C.}~\bibnamefont{Mazzoli}},
  \bibinfo{author}{\bibfnamefont{K.}~\bibnamefont{Kummer}},
  \bibinfo{author}{\bibfnamefont{G.}~\bibnamefont{Ghiringhelli}},
  \bibinfo{author}{\bibfnamefont{L.}~\bibnamefont{Braicovich}},
  \bibinfo{author}{\bibfnamefont{T.}~\bibnamefont{Tohyama}},
  \bibnamefont{et~al.}, \bibinfo{journal}{Nat. Commun.}
  \textbf{\bibinfo{volume}{5}}, \bibinfo{pages}{3714} (\bibinfo{year}{2014}).

\bibitem[{\citenamefont{Lee et~al.}(2014)\citenamefont{Lee, Lee, Nowadnick,
  Gerber, Tabis, Huang, Strocov, Motoyama, Yu, Moritz et~al.}}]{wslee14}
\bibinfo{author}{\bibfnamefont{W.~S.} \bibnamefont{Lee}},
  \bibinfo{author}{\bibfnamefont{J.~J.} \bibnamefont{Lee}},
  \bibinfo{author}{\bibfnamefont{E.~A.} \bibnamefont{Nowadnick}},
  \bibinfo{author}{\bibfnamefont{S.}~\bibnamefont{Gerber}},
  \bibinfo{author}{\bibfnamefont{W.}~\bibnamefont{Tabis}},
  \bibinfo{author}{\bibfnamefont{S.~W.} \bibnamefont{Huang}},
  \bibinfo{author}{\bibfnamefont{V.~N.} \bibnamefont{Strocov}},
  \bibinfo{author}{\bibfnamefont{E.~M.} \bibnamefont{Motoyama}},
  \bibinfo{author}{\bibfnamefont{G.}~\bibnamefont{Yu}},
  \bibinfo{author}{\bibfnamefont{B.}~\bibnamefont{Moritz}},
  \bibnamefont{et~al.}, \bibinfo{journal}{Nat. Phys.}
  \textbf{\bibinfo{volume}{10}}, \bibinfo{pages}{883} (\bibinfo{year}{2014}).

\bibitem[{\citenamefont{Yamase et~al.}(2015)\citenamefont{Yamase, Bejas, and
  Greco}}]{yamase15b}
\bibinfo{author}{\bibfnamefont{H.}~\bibnamefont{Yamase}},
  \bibinfo{author}{\bibfnamefont{M.}~\bibnamefont{Bejas}}, \bibnamefont{and}
  \bibinfo{author}{\bibfnamefont{A.}~\bibnamefont{Greco}},
  \bibinfo{journal}{Europhys. Lett.} \textbf{\bibinfo{volume}{111}},
  \bibinfo{pages}{57005} (\bibinfo{year}{2015}).

\bibitem[{\citenamefont{{A. Greco, H. Yamase, and M. Bejas}}(2016)}]{greco16}
\bibinfo{author}{\bibnamefont{{A. Greco, H. Yamase, and M. Bejas}}},
  \bibinfo{journal}{Phys. Rev. B} \textbf{\bibinfo{volume}{94}},
  \bibinfo{pages}{075139} (\bibinfo{year}{2016}).

\bibitem[{\citenamefont{Khaliullin and Horsch}(1996)}]{khaliullin96}
\bibinfo{author}{\bibfnamefont{G.}~\bibnamefont{Khaliullin}} \bibnamefont{and}
  \bibinfo{author}{\bibfnamefont{P.}~\bibnamefont{Horsch}},
  \bibinfo{journal}{Phys.\ Rev.\ B} \textbf{\bibinfo{volume}{54}},
  \bibinfo{pages}{R9600} (\bibinfo{year}{1996}).

\bibitem[{\citenamefont{Anderson}(1987)}]{anderson87}
\bibinfo{author}{\bibfnamefont{P.~W.} \bibnamefont{Anderson}},
  \bibinfo{journal}{Science} \textbf{\bibinfo{volume}{235}},
  \bibinfo{pages}{1196} (\bibinfo{year}{1987}).

\bibitem[{\citenamefont{{P. A. Lee, N. Nagaosa, and X.-G. Wen}}(2006)}]{lee06}
\bibinfo{author}{\bibnamefont{{P. A. Lee, N. Nagaosa, and X.-G. Wen}}},
  \bibinfo{journal}{Rev. Mod. Phys.} \textbf{\bibinfo{volume}{78}},
  \bibinfo{pages}{17} (\bibinfo{year}{2006}).

\bibitem[{\citenamefont{{A. Foussats and A. Greco}}(2004)}]{foussats04}
\bibinfo{author}{\bibnamefont{{A. Foussats and A. Greco}}},
  \bibinfo{journal}{Phys.\ Rev.\ B} \textbf{\bibinfo{volume}{70}},
  \bibinfo{pages}{205123} (\bibinfo{year}{2004}).

\bibitem[{\citenamefont{Bejas et~al.}(2012)\citenamefont{Bejas, Greco, and
  Yamase}}]{bejas12}
\bibinfo{author}{\bibfnamefont{M.}~\bibnamefont{Bejas}},
  \bibinfo{author}{\bibfnamefont{A.}~\bibnamefont{Greco}}, \bibnamefont{and}
  \bibinfo{author}{\bibfnamefont{H.}~\bibnamefont{Yamase}},
  \bibinfo{journal}{Phys.\ Rev.\ B} \textbf{\bibinfo{volume}{86}},
  \bibinfo{pages}{224509} (\bibinfo{year}{2012}).

\bibitem[{\citenamefont{Hoang and Thalmeier}(2002)}]{hoang02}
\bibinfo{author}{\bibfnamefont{A.~T.} \bibnamefont{Hoang}} \bibnamefont{and}
  \bibinfo{author}{\bibfnamefont{P.}~\bibnamefont{Thalmeier}},
  \bibinfo{journal}{J. Phys.: Condens. Matter} \textbf{\bibinfo{volume}{14}},
  \bibinfo{pages}{6639} (\bibinfo{year}{2002}).

\bibitem[{\citenamefont{{J. Merino, A. Greco, R.~H. McKenzie, and M.
  Calandra}}(2003)}]{merino03}
\bibinfo{author}{\bibnamefont{{J. Merino, A. Greco, R.~H. McKenzie, and M.
  Calandra}}}, \bibinfo{journal}{Phys.\ Rev.\ B} \textbf{\bibinfo{volume}{68}},
  \bibinfo{pages}{245121} (\bibinfo{year}{2003}).

\bibitem[{\citenamefont{Prelov\v{s}ek and Horsch}(1999)}]{prelovsek99}
\bibinfo{author}{\bibfnamefont{P.}~\bibnamefont{Prelov\v{s}ek}}
  \bibnamefont{and} \bibinfo{author}{\bibfnamefont{P.}~\bibnamefont{Horsch}},
  \bibinfo{journal}{Phys. Rev. B} \textbf{\bibinfo{volume}{60}},
  \bibinfo{pages}{R3735} (\bibinfo{year}{1999}).

\bibitem[{\citenamefont{{Z. Wang, Y. Bang, and G. Kotliar}}(1991)}]{wang91}
\bibinfo{author}{\bibnamefont{{Z. Wang, Y. Bang, and G. Kotliar}}},
  \bibinfo{journal}{Phys.\ Rev.\ Lett.} \textbf{\bibinfo{volume}{67}},
  \bibinfo{pages}{2733} (\bibinfo{year}{1991}).

\bibitem[{\citenamefont{Gehlhoff and Zeyher}(1995)}]{gehlhoff95}
\bibinfo{author}{\bibfnamefont{L.}~\bibnamefont{Gehlhoff}} \bibnamefont{and}
  \bibinfo{author}{\bibfnamefont{R.}~\bibnamefont{Zeyher}},
  \bibinfo{journal}{Phys.\ Rev.\ B} \textbf{\bibinfo{volume}{52}},
  \bibinfo{pages}{4635} (\bibinfo{year}{1995}).

\bibitem[{\citenamefont{Zeyher and Kuli\'c}(1996)}]{zeyher96a}
\bibinfo{author}{\bibfnamefont{R.}~\bibnamefont{Zeyher}} \bibnamefont{and}
  \bibinfo{author}{\bibfnamefont{M.}~\bibnamefont{Kuli\'c}},
  \bibinfo{journal}{Phys.\ Rev.\ B} \textbf{\bibinfo{volume}{54}},
  \bibinfo{pages}{8985} (\bibinfo{year}{1996}).

\bibitem[{\citenamefont{{A. Foussats and A. Greco}}(2002)}]{foussats02}
\bibinfo{author}{\bibnamefont{{A. Foussats and A. Greco}}},
  \bibinfo{journal}{Phys.\ Rev.\ B} \textbf{\bibinfo{volume}{65}},
  \bibinfo{pages}{195107} (\bibinfo{year}{2002}).

\bibitem[{\citenamefont{Emery and Kivelson}(1993)}]{emery93}
\bibinfo{author}{\bibfnamefont{V.~J.} \bibnamefont{Emery}} \bibnamefont{and}
  \bibinfo{author}{\bibfnamefont{S.~A.} \bibnamefont{Kivelson}},
  \bibinfo{journal}{Physica C} \textbf{\bibinfo{volume}{209}},
  \bibinfo{pages}{597} (\bibinfo{year}{1993}).

\bibitem[{\citenamefont{Tohyama and Maekawa}(1994)}]{tohyama94}
\bibinfo{author}{\bibfnamefont{T.}~\bibnamefont{Tohyama}} \bibnamefont{and}
  \bibinfo{author}{\bibfnamefont{S.}~\bibnamefont{Maekawa}},
  \bibinfo{journal}{Phys. Rev. B} \textbf{\bibinfo{volume}{49}},
  \bibinfo{pages}{3596} (\bibinfo{year}{1994}).

\bibitem[{\citenamefont{{E. Koch and R. Zeyher}}(2004)}]{koch04}
\bibinfo{author}{\bibnamefont{{E. Koch and R. Zeyher}}},
  \bibinfo{journal}{Phys.\ Rev.\ B} \textbf{\bibinfo{volume}{70}},
  \bibinfo{pages}{094510} (\bibinfo{year}{2004}).

\bibitem[{\citenamefont{Tohyama et~al.}(1995)\citenamefont{Tohyama, Horsch, and
  Maekawa}}]{tohyama95}
\bibinfo{author}{\bibfnamefont{T.}~\bibnamefont{Tohyama}},
  \bibinfo{author}{\bibfnamefont{P.}~\bibnamefont{Horsch}}, \bibnamefont{and}
  \bibinfo{author}{\bibfnamefont{S.}~\bibnamefont{Maekawa}},
  \bibinfo{journal}{Phys. Rev. Lett.} \textbf{\bibinfo{volume}{74}},
  \bibinfo{pages}{980} (\bibinfo{year}{1995}).

\end{thebibliography}

\end{document}